\documentclass[12pt]{article}
\usepackage{latexsym}
\usepackage{amsmath,amsfonts}
\usepackage{times}

\catcode`\@=11

\newcount\hour
\newcount\minute
\newtoks\amorpm \hour=\time\divide\hour by 60\minute
=\time{\multiply\hour by 60 \global\advance\minute by-\hour}
\edef\standardtime{{\ifnum\hour<12 \global\amorpm={am}%
        \else\global\amorpm={pm}\advance\hour by-12 \fi
        \ifnum\hour=0 \hour=12 \fi
        \number\hour:\ifnum\minute<10
        0\fi\number\minute\the\amorpm}}
\edef\militarytime{\number\hour:\ifnum\minute<10 0\fi\number\minute}

\def\draftlabel#1{{\@bsphack\if@filesw {\let\thepage\relax
   \xdef\@gtempa{\write\@auxout{\string
      \newlabel{#1}{{\@currentlabel}{\thepage}}}}}\@gtempa
   \if@nobreak \ifvmode\nobreak\fi\fi\fi\@esphack}
        \gdef\@eqnlabel{#1}}
\def\@eqnlabel{}
\def\@vacuum{}
\def\marginnote#1{}
\def\draftmarginnote#1{\marginpar{\raggedright\scriptsize\tt#1}}
\def\draft{
        \pagestyle{plain}
        \overfullrule=2pt
        \oddsidemargin -.5truein
        \def\@oddhead{\sl \phantom{\today\quad\militarytime} \hfil
        \smash{\Large\sl DRAFT} \hfil \today\quad\militarytime}
        \let\@evenhead\@oddhead
        \let\label=\draftlabel
        \let\marginnote=\draftmarginnote
        \def\ps@empty{\let\@mkboth\@gobbletwo
        \def\@oddfoot{\hfil \smash{\Large\sl DRAFT} \hfil}
        \let\@evenfoot\@oddhead}
        \def\@eqnnum{(\theequation)\rlap{\kern\marginparsep\tt\@eqnlabel}%
        \global\let\@eqnlabel\@vacuum}  }

\newcommand{\rf}[1]{(\ref{#1})}
\renewcommand{\theequation}{\thesection.\arabic{equation}}
\renewcommand{\thefootnote}{\fnsymbol{footnote}}
\newcommand{\newsection}{   
\setcounter{equation}{0}\section}

\def\appendix#1{\addtocounter{section}{1}\setcounter{equation}{0}
\renewcommand{\thesection}{\Alph{section}}
\section*{Appendix \thesection\protect\indent \parbox[t]{11.15cm}{#1}}
\addcontentsline{toc}{section}{Appendix \thesection\ \ \ #1}}

\def\be{\begin{equation}}
\def\ee{\end{equation}}
\def\beq{\begin{eqnarray}}
\def\eeq{\end{eqnarray}}

\def\parline{\,\partial\kern -0.55em /\,\,}

\def\half{{\frac{1}{2}}}

\def\AA{{\cal A}}

\def\EE{{\cal E}}

\def\LL{{\cal L}}
\def\MM{{\cal M}}
\def\NN{{\cal N}}

\def\TT{{\cal T}}

\def\Cbf{{\bf C}}
\def\Dbf{{\bf D}}
\def\Gbf{{\bf G}}
\def\Ibf{{\bf I}}
\def\ebf{{\bf e}}
\def\mbf{{\bf m}}
\def\alphabf{{\boldsymbol{\alpha}}}
\def\mubf{{\boldsymbol{\mu}}}
\def\Pibf{{\boldsymbol{\Pi}}}

\def\phik{|\phi\rangle}
\def\phibr{\langle\phi|}

\def\Phik{|\Phi\rangle}
\def\Phibr{\langle\Phi|}

\def\xik{|\xi\rangle}
\def\Xik{|\Xi\rangle}

\def\smzero{{\scriptscriptstyle (0)}}
\def\smone{{\scriptscriptstyle (1)}}
\def\smtwo{{\scriptscriptstyle (2)}}

\def\smx#1{{\scriptscriptstyle (#1)}}

\def\smzero{{\scriptscriptstyle (0)}}
\def\smone{{\scriptscriptstyle (1)}}
\def\smtwo{{\scriptscriptstyle (2)}}

\def\smx#1{{\scriptscriptstyle (#1)}}

\def\smpone{{\scriptscriptstyle [1]}}

\def\smponetwo{{\scriptscriptstyle [1,2]}}

\def\Iwt{\widetilde{I}}

\def\alpar{\alpha\partial}
\def\albpar{\bar\alpha\partial}

\def\Cb{\bar{C}}
\def\eb{\bar{e}}
\def\mb{\bar{m}}

\def\st{{\rm st}}
\def\total{{\rm total}}

\begin{document}


\begin{flushright}
FIAN-TD-2009-11 \qquad \ \ \ \ \  \\
arXiv: 0907.2207 [hep-th] \\
\end{flushright}

\vspace{1cm}

\begin{center}

{\Large \bf CFT adapted gauge invariant formulation of massive

\medskip
arbitrary spin fields in AdS }

\vspace{2.5cm}

R.R. Metsaev%
\footnote{ E-mail: metsaev@lpi.ru.
This work was supported by the RFBR Grant No.08-02-00963, RFBR Grant for
Leading Scientific Schools, Grant No. 1615.2008.2, by the Dynasty Foundation
and by the Alexander von Humboldt Foundation Grant PHYS0167.
}

\vspace{1cm}

{\it Department of Theoretical Physics, P.N. Lebedev Physical
Institute, \\ Leninsky prospect 53,  Moscow 119991, Russia }

\vspace{3.5cm}

{\bf Abstract}

\end{center}

Using Poincar\'e parametrization of AdS space, we study massive totally
symmetric arbitrary spin fields in AdS space of dimension greater than or
equal to four. CFT adapted gauge invariant formulation for such fields is
developed. Gauge symmetries are realized by using Stueckelberg formulation of
massive fields. We demonstrate that the mass parameter, curvature and radial
coordinate contributions to the gauge transformation and Lagrangian of the
AdS massive fields can be expressed in terms of ladder operators. Three
representations for the Lagrangian are discussed. Realization of the global
AdS symmetries in the conformal algebra basis is obtained. Modified de Donder
gauge leading to simple gauge fixed Lagrangian is found. The modified de
Donder gauge leads to decoupled equations of motion which can easily be
solved in terms of the Bessel function. New simple representation for gauge
invariant Lagrangian of massive (A)dS field in arbitrary coordinates is
obtained. Light-cone gauge Lagrangian of massive AdS field is also presented.

\newpage
\renewcommand{\thefootnote}{\arabic{footnote}}
\setcounter{footnote}{0}

\section{Introduction}

Further progress in understanding AdS/CFT correspondence requires, among
other things, better understanding of field dynamics in $AdS$ space. Although
many interesting approaches to $AdS$ fields are known in the literature (for
review see Refs.\cite{Bekaert:2005vh}-\cite{Fotopoulos:2008ka}), analysis of
concrete dynamical aspects of such fields is still a challenging procedure.
One of ways to simplify analysis of field and string dynamics in $AdS$ space
is based on use of the Poincar\'e parametrization of $AdS$ space%
\footnote{ Studying $AdS_5\times S^5$ superstring action
\cite{Metsaev:1998it} in Poincar\'e parametrization may be found in
Ref.\cite{Metsaev:2000ds}. Recent interesting application of Poincar\'e
coordinates to studying $AdS_5\times S^5$ string $T-$duality  may be found in
Refs.\cite{Beisert:2008iq,Berkovits:2008ic}.}.
Use of the Poincar\'e coordinates simplifies analysis of many aspect of $AdS$
field dynamics and therefore these coordinates have extensively been used for
studying the AdS/CFT correspondence. In Ref.\cite{Metsaev:2008ks}, we
developed a approach which  is based on considering of $AdS$ field dynamics
in the Poincar\'e coordinates and applied our approach to study of massless
$AdS$ fields. We think that our approach might be useful for study of $AdS$
string massive modes. Therefore it is desirable to generalize our approach to
the case of massive $AdS$ fields. This is that what we do in this paper.
Namely, using the Poincar\'e parametrization of $AdS$ space we discuss
massive totally symmetric arbitrary spin-$s$, $s\geq 1$, bosonic field
propagating in $AdS_{d+1}$ space of dimension $d+1 \geq 4$. Our results can
be summarized as follows.

i) Using the Poincar\'e parametrization of $AdS$, we obtain gauge invariant
Lagrangian for free massive arbitrary spin $AdS$ field.  The Lagrangian is
{\it explicitly invariant with respect to boundary Poincar\'e symmetries},
i.e., manifest symmetries of our Lagrangian are adapted to manifest
symmetries of boundary CFT. We show that all the mass parameter, curvature
and radial coordinate contributions to our Lagrangian and gauge
transformation are entirely expressed in terms of ladder operators that
depend on the mass parameter, radial coordinate and radial derivative.
General structure of the Lagrangian we use is the same as the one for
massless $AdS$ fields. Lagrangian of massive $AdS$ field is distinguished by
appropriate ladder operators. We find two new concise expressions for the
gauge invariant Lagrangian.

ii) We generalize modified de Donder gauge, found for massless $AdS$ fields
in Ref.\cite{Metsaev:2008ks}, to the case of massive fields. As in the case
of massless fields, the modified de Donder gauge leads to simple gauge
fixed Lagrangian and {\it decoupled equations of motion}%
\footnote{ Our modified de Donder gauge seems to be unique gauge that leads
to decoupled equations of motion. Light-cone gauge \cite{Metsaev:1999ui} also
leads to decoupled equations of motion, but the light-cone gauge breaks
boundary Lorentz symmetries.}.
Note that the standard de Donder gauge leads to coupled equations of motion
whose solutions for $s\geq 2$ are not known in closed form even for massless
$AdS$ fields. In contrast to this, our modified de Donder gauge leads to
simple decoupled equations which are easily solved in terms of the Bessel
function.

\newsection{ Lagrangian and its gauge and global symmetries}

We begin with discussion of field content of our approach. In
Ref.\cite{Zinoviev:2001dt}, the massive spin-$s$ field propagating in
$AdS_{d+1}$ space is described by double-traceless $so(d,1)$
algebra totally symmetric tensor fields $\Phi^{A_1\ldots A_{s'}}$,
$s'=0,1,\ldots,s$.%
\footnote{ $A,B,C=0,1,\ldots, d$ and $a,b,c=0,1,\ldots, d-1$ are the
respective flat vector indices of the $so(d,1)$ and $so(d-1,1)$ algebras. In
Poincar\'e parametrization of $AdS_{d+1}$ space, $ds^2=(dx^adx^a +
dzdz)/z^2$. We use the conventions: $\partial_a \equiv\partial/\partial x^a$,
$\partial_z\equiv\partial/\partial z$. Vectors of $so(d,1)$ algebra are
decomposed as $X^A=(X^a,X^z)$.}
These tensor fields can be decomposed in real-valued scalar, vector, and
totally symmetric tensor fields of the $so(d-1,1)$ algebra:
\be \label{18052008-02}
\phi_\lambda^{a_1\ldots a_{s'}}\,, \hspace{2cm} \lambda=[s-s']_2\,,\qquad
s'=0,1,\ldots,s-1,s\,.
\ee
Henceforth, the notation $\lambda=[n]_2$ implies that
$\lambda=-n,-n+2,-n+4,\ldots,n-4, n-2,n$. To illustrate the field content
given in \rf{18052008-02} we use shortcut $\phi_{(s',\lambda)}$ for the field
$\phi_\lambda^{a_1\ldots a_{s'}}$ and note that fields in \rf{18052008-02}
can be represented as
{\small
\be
\begin{array}{ccccccccc}
&  & &  & \phi_{(s,0)}& & & &
\\[12pt]
&  & & \phi_{(s-1,-1)}  & & \phi_{(s-1,1)} & & &
\\[12pt]
& & \ldots  &   & \ldots & & \ldots  &  &
\\[12pt]
& \phi_{(1,1-s)} & & \phi_{(1,3-s)} & \ldots & \phi_{(1,s-3)} & &
\phi_{(1,s-1)} &
\\[12pt]
\phi_{(0,-s)} & & \phi_{(0,2-s)} & & \ldots & & \phi_{(0,s-2)} & &
\phi_{(0,s)}
\end{array}
\ee
}
The fields $\phi_\lambda^{a_1\ldots a_{s'}}$ with $s' \geq 4$
are double-traceless,%
\footnote{ Note that $so(d-1,1)$ tensorial components of the
Fronsdal-Zinoviev fields $\Phi^{A_1\ldots A_{s'}}$ are not double-traceless.
Using appropriate transformation (see \rf{05012008}) those tensorial
components can be transformed to our fields in \rf{18052008-02}.}
\be \label{18052008-03} \phi_\lambda^{aabba_5\ldots a_{s'}}=0\,, \hspace{2cm}
\lambda=[s-s']_2\,, \qquad s'=4,5,\ldots,s. \ee
The fields in \rf{18052008-02} subject to constraints \rf{18052008-03}
constitute a field content of our approach.

To simplify presentation we use creation operators $\alpha^a$, $\alpha^z$,
$\zeta$ and the respective annihilation operators, $\bar{\alpha}^a$,
$\bar{\alpha}^z$, $\bar\zeta$. Then, fields \rf{18052008-02} can be collected
into a ket-vector $|\phi\rangle$
defined by%
\footnote{ We use oscillator formulation to handle the many indices appearing
for tensor fields (for review see
Refs.\cite{Bekaert:2006ix,Boulanger:2008up}). In a proper way, oscillators
arise in the framework of world-line approach to higher-spin fields (see e.g.
Refs.\cite{Bastianelli:2007pv,Cherney:2009mf}).}
\beq \label{01072009-01}
&& \phik = \sum_{s'=0}^s\,\,\sum_{\lambda=[s-s']_2}
\frac{\zeta_{\phantom{z}}^{\frac{s-s'+\lambda}{2}}
\alpha_z^{\frac{s-s'-\lambda}{2}}\alpha^{a_1}\ldots
\alpha^{a_{s'}}}{s'!\sqrt{(\frac{s-s'+\lambda}{2})!
(\frac{s-s'-\lambda}{2})!}} \, \phi_\lambda^{a_1\ldots a_{s'}} |0\rangle\,.
\eeq
From \rf{01072009-01} we see that the ket-vector $|\phi\rangle$ is degree-$s$
homogeneous polynomial in $\alpha^a$,
$\alpha^z$, $\zeta$.%
\footnote{ Throughout this paper we use the following notation for operators
constructed out the oscillators and derivatives: $N_\alpha \equiv \alpha^a
\bar\alpha^a$, $N_z \equiv \alpha^z \bar\alpha^z$, $N_\zeta \equiv \zeta
\bar\zeta$, $\alpha^2 =\alpha^a \alpha^a$, $\bar\alpha^2 = \bar\alpha^a
\bar\alpha^a$, $\Box=\partial^a\partial^a$, $\alpha\partial
=\alpha^a\partial^a$, $\bar\alpha\partial =\bar\alpha^a\partial^a$.} In terms
of the ket-vector $\phik$, double-tracelessness constraint \rf{18052008-03}
takes the form%
\footnote{ We adapt the formulation in terms of the double-traceless gauge
fields \cite{Fronsdal:1978vb} (see also
Refs.\cite{Zinoviev:2001dt,Metsaev:2006zy}). Discussion of various
formulations in terms of unconstrained gauge fields may be found in
Refs.\cite{Francia:2002aa}-\cite{Buchbinder:2007ak}. Study of other
interesting approaches which seem to be most suitable for the theory of
interacting fields may be found e.g. in
Refs.\cite{Alkalaev:2003qv,Iazeolla:2008ix}.}
\be
\label{phidoutracona01} (\bar{\alpha}^2)^2 \phik = 0 \,.
\ee

Action and Lagrangian we found take the form
\be \label{spi2lag01}  S = \int d^dx dz \ \LL\,,\qquad  \LL = \frac{1}{2}
\phibr E \phik\,, \ee
$\langle\phi| \equiv (\phik)^\dagger$. We now discuss various representations
for operator $E$ and the Lagrangian in turn.

{\it 1st representation}. This representation is given by
\beq
\label{Frosecordope01} E  & = & E_\smtwo + E_\smone + E_\smzero\,,
\\
\label{Frosecordope02} && E_\smtwo \equiv \Box -
\alpha\partial\bar\alpha\partial + \frac{1}{2}(\alpha\partial)^2\bar\alpha^2
+ \frac{1}{2} \alpha^2 (\bar\alpha\partial)^2 - \frac{1}{2}\alpha^2 \Box
\bar\alpha^2
-\frac{1}{4}\alpha^2\alpha\partial\,\bar\alpha\partial\bar\alpha^2\,,\qquad
\\
&& E_\smone \equiv  \eb_1 \AA + e_1 \bar\AA \,,
\\[3pt]
&& E_\smzero \equiv m_1 + \alpha^2\bar\alpha^2m_2 + \mb_3 \alpha^2 + m_3
\bar\alpha^2\,,
\eeq\beq
&&\hspace{1cm} \AA \equiv  \alpar - \alpha^2 \albpar +
\frac{1}{4}\alpha^2\,\alpar\,\bar\alpha^2 \,,
\\
\label{Abdef02} && \hspace{1cm} \bar\AA \equiv \albpar -  \alpar\bar\alpha^2
+ \frac{1}{4}\alpha^2\,\albpar\,\bar\alpha^2 \,,
\\[5pt]
&&\hspace{1cm} m_1  =   \eb_1 e_1 - 2\frac{2s + d-3 -2N_z-2N_\zeta}{2s + d-4
- 2N_z -2N_\zeta } e_1 \eb_1\,,
\\
&& \hspace{1cm} m_2  =   - \half \eb_1 e_1 +  \frac{1}{4} \frac{2s + d -2N_z
- 2N_\zeta}{2s + d-4 - 2N_z -2N_\zeta} e_1 \eb_1\,,
\\
\label{01072009-04} &&\hspace{1cm} m_3 = \frac{1}{2}e_1 e_1 \,,
\qquad \qquad
\mb_3  =  \frac{1}{2} \eb_1 \eb_1 \,,
\eeq
\beq
\label{e1def01} && - e_1 = \zeta r_\zeta \TT_{ -\nu - \half} + \alpha^z r_z
\TT_{\nu-\half} \,,
\qquad
- \eb_1 = \TT_{\nu + \half}  r_\zeta \bar\zeta  + \TT_{-\nu + \half} r_z
\bar\alpha^z \,,
\\[7pt]
\label{03072009-01} && \TT_\nu = \partial_z+ \frac{\nu}{z}\,,
\qquad
\nu = \kappa + N_\zeta - N_z\,,\qquad \kappa \equiv E_0 - \frac{d}{2}\,,
\\[7pt]
\label{01072009-05} && r_\zeta = \left(\frac{(s+\frac{d-4}{2}
-N_\zeta)(\kappa - s-\frac{d-4}{2} + N_\zeta)(\kappa + 1 +
N_\zeta)}{2(s+\frac{d-4}{2}-N_\zeta - N_z)(\kappa +N_\zeta -N_z) (\kappa+
N_\zeta - N_z +1)}\right)^{1/2}\,,
\\[7pt]
\label{01072009-06} && r_z = \left(\frac{(s+\frac{d-4}{2} -N_z)(\kappa + s +
\frac{d-4}{2} - N_z)(\kappa - 1 - N_z)}{2(s+\frac{d-4}{2}-N_\zeta-N_z)(\kappa
+ N_\zeta - N_z) (\kappa +N_\zeta - N_z -1)}\right)^{1/2}\,,
\eeq
where subscript $n$ in $E_{\smx{n}}$ \rf{Frosecordope01} tells us that
$E_{\smx{n}}$ is degree-$n$ homogeneous polynomial in the flat derivative
$\partial^a$. The following remarks are in order.
\\
i) The parameter $\kappa$ \rf{03072009-01}  is expressed in terms of spin-$s$
massive field lowest energy $E_0$. Using result in Ref.\cite{Metsaev:2003cu}
we can express $\kappa$ in terms of the standard mass parameter $m$,
\be \label{08072009-01} \kappa = \sqrt{m^2
+\Bigl(s+\frac{d-4}{2}\Bigr)^2}\,.\ee
ii) Operator $E_\smtwo$ \rf{Frosecordope02} is the symmetrized Fronsdal
operator represented in terms of the oscillators. This operator takes the
same form as the one of massless field in $d$-dimensional flat space. Thus,
the operator $E$ \rf{Frosecordope01} is given by the sum of the standard
Fronsdal operator $E_\smtwo$ and new operators $E_\smone$, $E_\smzero$ which
depend on the mass parameter $m$, the radial coordinate and derivative, $z$,
$\partial_z$.
\\
iii) Dependence of $E$ \rf{Frosecordope01} on the mass parameter $m$, the
radial coordinate and derivative, $z$, $\partial_z$, is entirely governed by
the operators $e_1$ and $\eb_1$ \rf{e1def01} which we will refer to as ladder
operators%
\footnote{ Interesting application of other ladder operators to studying
AdS/QCD correspondence may be found in Ref.\cite{Brodsky:2008pg}. We believe
that our approach will also be useful for better understanding of various
aspects of AdS/QCD correspondence which are discussed e.g. in
Refs.\cite{Brodsky:2008pg,Andreev:2002aw}.}.
\\
iv) Representation for the Lagrangian in \rf{spi2lag01} -\rf{01072009-04} is
universal and is valid for arbitrary Poincar\'e invariant theory. Various
Poincar\'e invariant theories are distinguished by ladder operators entering
the operator $E$. This is to say that the operators $E$ of massive and
conformal fields in flat space depend on the oscillators $\alpha^a$,
$\bar\alpha^a$ and the flat derivative $\partial^a$ in the same way as the
operator $E$ of $AdS$ fields \rf{Frosecordope01}. In other words, the
operators $E$ for massless and massive $AdS$ fields, massive and conformal
fields in flat space are distinguished only by the operators $e_1$ and
$\eb_1$. We note that it is finding the ladder operators that provides real
difficulty. Expressions for $e_1$, $\eb_1$ appropriate for conformal and
massive fields in flat space may be found in
Refs.\cite{Metsaev:2007fq,Metsaev:2008fs}.

{\it 2nd representation for the operator $E$}. Lagrangian can be presented in
the form given in \rf{spi2lag01} with the following concise expression for
the operator $E$:
\beq \label{29012009-01}
E & = & \mubf (\Box - \MM_\nu^2) - C\bar{C}\,,
\\[5pt]
\label{01072009-07} && \MM_\nu^2 \equiv - \partial_z^2 + \frac{1}{z^2}
(\nu^2-\frac{1}{4})\,,
\\[5pt]
\label{01072009-08} && \Cb \equiv  \albpar - \half \alpar \bar\alpha^2  -
\eb_1\Pi^\smponetwo + \half e_1 \bar\alpha^2\,,
\\[5pt]
\label{01072009-09} && C \equiv \alpar - \half \alpha^2 \albpar  - e_1
\Pi^\smponetwo + \half \eb_1 \alpha^2\,,
\\[5pt]
\label{18052008-08} && \mubf \equiv 1-
\frac{1}{4}\alpha^2\bar\alpha^2\,,\qquad  \Pi^\smponetwo \equiv 1
-\alpha^2\frac{1}{2(2N_\alpha +d)}\bar\alpha^2\,,
\eeq
where $\nu$ is given in \rf{03072009-01}. Operator $E$ in \rf{29012009-01}
differs from the one in \rf{Frosecordope01} by terms proportional to
$(\alpha^2)^2$ and $(\bar\alpha^2)^2$. Therefore, in view of
double-tracelessness constraint \rf{phidoutracona01}, these two
representations for $E$ lead to the same Lagrangian \rf{spi2lag01}. We note
that operator $E$ in \rf{29012009-01} respects, in contrast to the one in
\rf{Frosecordope01}, double-tracelessness constraint \rf{phidoutracona01}.
For massless field in $d$-dimensional {\it flat} space, $e_1=\eb_1=0$,
operator $\Cb$ \rf{01072009-08} coincides with the standard de Donder
operator. In terms of the ladder operators, mass operator $\MM^2$ takes the
form
\be \label{MMdef01} \MM^2 \equiv - \eb_1 e_1 + \frac{2s + d - 2 - 2N_z -
2N_\zeta}{2s + d - 4 - 2N_z-2N_\zeta} e_1 \eb_1\,, \ee
while the $\MM_\nu^2$ \rf{01072009-07} is obtained form \rf{MMdef01} by using
ladder operators given in \rf{e1def01}.

{\it CFT adapted representation of the Lagrangian}. Taking into account
representation for operator $E$ in \rf{29012009-01} and noticing the
relations
$\MM_\nu^2 = \TT_{\nu-\half}^\dagger \TT_{\nu-\half}$, $C = - \Cb^\dagger$,
where $\TT_\nu$ is given in \rf{03072009-01}, we see that Lagrangian
\rf{spi2lag01} can be represented as (up to total derivatives)
\be
\LL =  - \half \langle \partial^a \phi|\mubf | \partial^a \phi\rangle -\half
\langle \TT_{\nu-\half} \phi| \mubf | \TT_{\nu-\half} \phi\rangle + \half
\langle \bar{C}\phi|| \bar{C}\phi\rangle\,.
\ee
This form of the Lagrangian turns out to be very convenient for studying
AdS/CFT correspondence.

{\it Gauge symmetries}. We now discuss gauge symmetries of Lagrangian
\rf{spi2lag01}. To this end we introduce the following set of gauge
transformation parameters:
\be \label{epsilonset01}
\xi_\lambda^{a_1\ldots a_{s'}}\,,\qquad\qquad \lambda = [s-1-s']_2\,,\qquad
s'=0,1,\ldots,s-1\,. \ee
The gauge parameters $\xi_\lambda$, $\xi_\lambda^a$, and
$\xi_\lambda^{a_1\ldots a_{s'}}$, $s'\geq 2$ in \rf{epsilonset01}, are the
respective scalar, vector, and rank-$s'$ totally symmetric tensor fields of
the $so(d-1,1)$ algebra. The gauge parameters $\xi_\lambda^{a_1\ldots
a_{s'}}$ with $s'\geq 2 $ are subjected to the tracelessness constraint,
\be \label{epsdoutracon01} \xi_\lambda^{aaa_3\ldots a_{s'}}=0\,, \qquad\qquad
\lambda = [s-1-s']_2\,, \qquad s'= 2,3,\ldots, s-1\,. \ee
We now, as usually, collect gauge transformation parameters in ket-vector
$\xik$ defined by
\be
\xik = \sum_{s'=0}^{s-1} \,\,\,\sum_{\lambda=[s-1-s']_2}
\frac{\zeta_{\phantom{z}}^{\frac{s-1-s'+\lambda}{2}}
\alpha_z^{\frac{s-1-s'-\lambda}{2}}\alpha^{a_1}\ldots
\alpha^{a_{s'}}}{s'!\sqrt{(\frac{s-1-s'+\lambda}{2})!
(\frac{s-1-s'-\lambda}{2})!}} \, \xi_\lambda^{a_1\ldots a_{s'}} |0\rangle\,.
\ee
We note that the ket-vector $\xik$ is a degree-$(s-1)$ homogeneous polynomial
in the oscillators $\alpha^a$, $\alpha^z$, $\zeta$. In terms of the
ket-vector $\xik$, tracelessness constraint \rf{epsdoutracon01} takes the
form
\be \label{xialgcon01} \bar\alpha^2 \xik=0 \,.\ee
Lagrangian \rf{spi2lag01} is invariant under the following gauge
transformation:
\be \label{gautraarbspi01}
\delta \phik  = G\xik\,, \qquad G= \alpar - e_1 - \alpha^2\frac{1}{2N_\alpha
+ d- 2}\eb_1 \,,
\ee
where $e_1$, $\eb_1$ are given in \rf{e1def01}. From \rf{gautraarbspi01}, we
see that the mass parameter, radial coordinate and derivative contributions
to gauge transformation \rf{gautraarbspi01} are entirely expressed in terms
of the ladder operators $e_1$ and $\eb_1$. We note that use of operator $G$
\rf{gautraarbspi01} allows us to write new representation for the operator
$E$ entering Lagrangian \rf{spi2lag01},
\be \label{E3rddef01} E = \mubf (\Box - \MM_\nu^2 - G\Cb) \,.\ee

{\it Global $so(d,2)$ symmetries.} Relativistic symmetries of $AdS_{d+1}$
space are described by the $so(d,2)$ algebra. In our approach, the massive
spin-$s$ $AdS_{d+1}$ field is described by the set of the $so(d-1,1)$ algebra
fields \rf{18052008-02}. Therefore it is reasonable to represent the
$so(d,2)$ algebra so that to respect manifest $so(d-1,1)$ symmetries. For
application to the AdS/CFT correspondence, most convenient form of the
$so(d,2)$ algebra that respects the manifest $so(d-1,1)$ symmetries is
provided by nomenclature of the conformal algebra. This is to say that the
$so(d,2)$ algebra consists of translation generators $P^a$, conformal boost
generators $K^a$, dilatation generator $D$, and generators $J^{ab}$ which
span $so(d-1,1)$ algebra. Normalization for commutators of the $so(d,2)$
algebra generators we use may be found in formulas (3.1)-(3.4) in
Ref.\cite{Metsaev:2008ks}.

Requiring $so(d,2)$ symmetries implies that the action is invariant with
respect to transformation $\delta_{\hat{G}} \phik  = \hat{G} \phik$, where
the realization of $so(d,2)$ algebra generators $\hat{G}$ in terms of
differential operators acting on the ket-vector $\phik$ takes the form
\beq
\label{conalggenlis01} && P^a = \partial^a \,,
\qquad
J^{ab} = x^a\partial^b -  x^b\partial^a + M^{ab}\,,
\\[3pt]
\label{conalggenlis03} && D = x\partial  + \Delta\,,
\qquad
\Delta \equiv  z\partial_z + \frac{d-1}{2}\,,
\\[3pt]
\label{conalggenlis04} && K^a = -\frac{1}{2}x^2\partial^a + x^a D + M^{ab}x^b
+ R^a \,,
\eeq
$x\partial\equiv x^a\partial^a$, $x^2\equiv x^ax^a$. In
\rf{conalggenlis01},\rf{conalggenlis04}, $M^{ab}$ is spin operator of the
$so(d-1,1)$ algebra. Representation of $M^{ab}$ and operator $R^a$
\rf{conalggenlis04} on space of ket-vector $\phik$ \rf{01072009-01} takes the
form
\beq
&& M^{ab} = \alpha^a \bar\alpha^b - \alpha^b \bar\alpha^a\,,
\\[5pt]
&& R^a = z \Iwt^a ( r_\zeta \bar\zeta + r_z \bar\alpha^z) - z (\zeta r_\zeta
+\alpha^z r_z )\bar\alpha^a - \half z^2 \partial^a\,,
\\[5pt]
&& \hspace{1cm} \Iwt^a \equiv \alpha^a - \alpha^2 \frac{1}{2N_\alpha + d
-2}\bar\alpha^a\,,
\eeq
where $r_\zeta$, $r_z$ are given in \rf{01072009-05},\rf{01072009-06}. We see
that realization of Poincar\'e symmetries on bulk $AdS$ fields
\rf{conalggenlis01} coincide with realization of Poincar\'e symmetries on
boundary CFT operators. Note that realization of $D$- and $K^a$-symmetries on
bulk $AdS$ fields \rf{conalggenlis03},\rf{conalggenlis04} coincides, by
module of contributions of operators $\Delta$ and $R^a$, with the realization
of $D$- and $K^a$-symmetries on boundary CFT operators. Realizations of the
$so(d,2)$ algebra on bulk $AdS$ fields and boundary CFT operators are
distinguished by $\Delta$ and $R^a$. The realization of the $so(d,2)$
symmetries given in \rf{conalggenlis01}-\rf{conalggenlis04} turns out to be
very convenient for studying AdS/CFT correspondence \cite{Metsaev:2008fs}.

\newsection{  Modified de Donder gauge}

To discuss modified de Donder gauge we use representation for Lagrangian
given in \rf{spi2lag01},\rf{29012009-01}. It is easy to see that use of the
following {\it modified de Donder} gauge-fixing term
\beq
\label{22082008-01} \LL_{{\rm g.fix}} & = & \half \phibr  C\bar{C} \phik \,,
\eeq
leads to the surprisingly simple gauge fixed Lagrangian $\LL_\total$,
$\LL_\total \equiv \LL + \LL_{{\rm g.fix}}$,
\beq \label{20082008-05}
\label{20082008-06} && \LL_\total = \half \phibr E_\total \phik \,,
\qquad
E_\total =  (1-\frac{1}{4}\alpha^2 \bar\alpha^2) (\Box - \MM_\nu^2)\,,
\eeq
where $\MM_\nu^2$ is given in \rf{01072009-07}. We note that our gauge-fixing
term \rf{22082008-01} respects the Poincar\'e and dilatation symmetries but
breaks the conformal boost $K^a$-symmetries, i.e., the simple form of gauge
fixed Lagrangian \rf{20082008-05} is achieved at the cost of the
$K^a$-symmetries. In terms of tensorial components, gauge fixed Lagrangian
\rf{20082008-05} takes the form
\beq
\label{02072009-01} &&  \LL_\total = \sum_{s'=0}^s\,\, \sum_{\lambda =
[s-s']_2} \LL_{s',\lambda}\,,
\\[7pt]
\label{02072009-02}&&  \LL_{s',\lambda} = \frac{1}{2 s'!}\Bigl(
\phi_\lambda^{a_1\ldots a_{s'}} \Box_{\kappa+\lambda}^{}
\phi_\lambda^{a_1\ldots a_{s'}} - \frac{s'(s'-1)}{4}
\phi_\lambda^{aaa_3\ldots a_{s'}} \Box_{\kappa+\lambda}^{}
\phi_\lambda^{bba_3\ldots a_{s'}}\Bigr)\,,
\\[7pt]
\label{02072009-03} && \hspace{1cm} \Box_{\kappa+\lambda} \equiv \Box
+\partial_z^2 - \frac{1}{z^2}\Bigl( (\kappa + \lambda)^2
-\frac{1}{4}\Bigr)\,.
\eeq
We see that the modified de Donder gauge fixing leads to simple gauge fixed
Lagrangian.

We now discuss gauge-fixing procedure at the level of equations of motion.
Representation for the operator $E$ in \rf{E3rddef01} turns out to be
convenient for this purpose.%
\footnote{ Operators $E$ in \rf{29012009-01},\rf{E3rddef01} respect, in
contrast to operator $E$ in \rf{Frosecordope01}, double-tracelessness
constraint \rf{phidoutracona01}.}
This is to say that Lagrangian with $E$ in \rf{E3rddef01} leads to the
following gauge invariant equations of motion
\be \label{01072009-10} (\Box - \MM_\nu^2 - G\Cb) \phik = 0\,. \ee
{\it Modified de Donder gauge} condition is then defined to be
\be \label{080405-01} \bar{C} \phik = 0\,, \ee
where $\bar{C}$ is given in \rf{01072009-08}. The fact that this gauge is
accessible with gauge transformation \rf{gautraarbspi01} can be proved as
follows. i) By virtue of \rf{phidoutracona01}, we have the relation
$\bar\alpha^2\bar{C}\phik=0$ which implies that gauge condition
\rf{080405-01} respects constraint for gauge transformation parameter $\xik$,
\rf{xialgcon01}; ii) Gauge variation of $\bar{C} \phik$ is given by $\delta
(\bar{C} \phik) = (\Box-\MM_\nu^2)\xik$. Making standard assumption that the
operator $\Box-\MM_\nu^2$ is invertible, we see that gauge condition
\rf{080405-01} is indeed accessible.

Using the modified de Donder gauge condition in gauge invariant equations of
motion \rf{01072009-10} leads to the following gauge fixed equations of
motion:
\be \label{20082008-03} (\Box - \MM_\nu^2 )\phik = 0 \,, \ee
where $\MM_\nu^2$ is defined in \rf{01072009-07}. In terms of fields
\rf{18052008-02}, equations \rf{20082008-03} can be represented as
\be \label{23082008-01}  \Box_{\kappa+\lambda}^{} \phi_\lambda^{a_1\ldots
a_{s'}} = 0\,,\qquad
\lambda=[s-s']_2\,,\qquad s'=0,1,\ldots, s\,,
\ee
where $\Box_{\kappa+\lambda}$ is given in \rf{02072009-03}. Thus, our {\it
modified de Donder gauge condition \rf{080405-01} leads to decoupled
equations of motion} \rf{23082008-01} which can easily be solved in terms of
the Bessel function
\footnote{ Interesting method of solving $AdS$ field equations of motion
which is based on star algebra products in auxiliary spinor variables is
discussed in Refs.\cite{Bolotin:1999fa,Didenko:2009td}. As a side of remark
we note that our modified de Donder gauge can be generalized to conformal
flat spaces (see Appendix D in Ref.\cite{Metsaev:2008fs}.)}.
For spin-1 field, gauge condition \rf{080405-01} turns out to be a
modification of the Lorentz gauge.

\newsection{ Comparison of standard and modified de Donder gauges}
\label{comparissec}

Our approach to the massive spin-$s$ field in $AdS_{d+1}$ is based on use of
double-traceless $so(d-1,1)$ algebra fields \rf{18052008-02}. One of popular
approaches to the massive spin-$s$ field in $AdS_{d+1}$ is based on use of
double-traceless $so(d,1)$ algebra fields $\Phi^{A_1\ldots A_{s'}}$,
$s'=0,1,\ldots,s$, \cite{Zinoviev:2001dt}. In this Section, our aims are as
follows. i) Using the fields $\Phi^{A_1\ldots A_{s'}}$ and arbitrary
parametrization of $(A)dS$ space, we find new representation for gauge
invariant Lagrangian of massive $(A)dS$ field and standard de Donder gauge
condition;
\footnote{ To our knowledge,  the standard de Donder gauge for arbitrary spin
massive $(A)dS$ fields has not been discussed in earlier literature. Study of
the standard de Donder gauge for flat arbitrary spin massive fields may be
found in Ref.\cite{Metsaev:2008fs}. Recent applications of the {\it standard}
de Donder gauge to the various problems of massless fields may be found in
Refs.\cite{Guttenberg:2008qe,Manvelyan:2008ks}.}
ii) We explain how our modified de Donder gauge \rf{080405-01} is represented
in terms of the fields $\Phi^{A_1\ldots A_{s'}}$. iii) We show explicitly how
our fields \rf{18052008-02} are related to the fields $\Phi^{A_1\ldots
A_{s'}}$.

{\it New representation for gauge invariant Lagrangian of massive field in
$(A)dS_{d+1}$}. We begin with discussion of gauge invariant Lagrangian using
arbitrary coordinates of $(A)dS$. To simplify the presentation we introduce
ket-vector $\Phik$,
\beq
\label{froketvect} && \Phik \equiv \sum_{s'=0}^s
\frac{\zeta^{s-s'} \alpha^{A_1} \ldots \alpha^{A_{s'}} }{s'!\sqrt{(s - s')!}}
\, \Phi^{A_1\ldots A_{s'}} |0\rangle\,,
\\
\label{froketvectcon} && (\bar\alphabf^2)^2 |\Phi\rangle=0,
\eeq
$\alphabf^2 \equiv \alpha^A\alpha^A$, $\bar\alphabf^2 \equiv
\bar\alpha^A\bar\alpha^A$, where \rf{froketvectcon} tells us that the
$\Phi^{A_1\ldots A_{s'}}$ are double-traceless. We find the following concise
expression
for gauge invariant Lagrangian of massive spin-$s$ field in $(A)dS_{d+1}$:%
\beq
\label{0207200-04}\LL & = & \half e \langle \Phi| E \Phik \,,
\\[5pt]
\label{0207200-05} E & = & (1-\frac{1}{4}\alphabf^2 \bar\alphabf^2)
(\Box_{_{\rm (A)dS}} + \mbf_1 + \rho \alphabf^2\bar\alphabf^2) - \Cbf_\st
\bar{\Cbf}_\st\,,
\\[5pt]
\label{0207200-06} && \mbf_1 = -m^2 +\rho \Bigl( s(s+d-5) -2d+ 4
+N_\zeta(2s+d-1-N_\zeta)\Bigr)\,,
\\[5pt]
\label{0207200-07} && \bar\Cbf_\st \equiv  \bar\alphabf \Dbf - \half \alphabf
\Dbf  \bar\alphabf^2 - \bar\ebf_1\Pibf^\smponetwo + \half \ebf_1
\bar\alphabf^2\,,
\\[5pt]
\label{0207200-08} && \Cbf_\st \equiv \alphabf \Dbf  - \half \alphabf^2
\bar\alphabf \Dbf  - \ebf_1 \Pibf^\smponetwo + \half \bar\ebf_1 \alphabf^2\,,
\\
&& \Pibf^\smponetwo = 1 - \alphabf^2\frac{1}{2(2N_\alphabf + d +
1)}\bar\alphabf^2\,,\qquad N_\alphabf \equiv \alpha^A\bar\alpha^A\,,
\eeq

\be
\ebf_1 = \zeta \widetilde\ebf_1\,, \qquad
\bar\ebf_1 = -\widetilde\ebf_1 \bar\zeta\,,
\ee

\be \label{09072009-01}
\widetilde\ebf_1
=\Bigl(\frac{2s+d-3-N_\zeta}{2s+d-3-2N_\zeta}\Bigr)^{1/2}\Bigl(m^2 -\rho
N_\zeta (2s+d-4-N_\zeta)\Bigr)^{1/2}\,,
\ee

\beq
&& \Box_{_{\rm (A)dS}}\equiv D^A D^A + \omega^{AAB}D^B\,,
\qquad
\bar\alphabf {\bf D} \equiv \bar\alpha^A D^A\,,
\qquad
\alphabf {\bf D} \equiv \alpha^A D^A\,,
\eeq
where $e=\det e_\mu^A$, $e_\mu^A$ stands for vielbein of $(A)dS_{d+1}$ space,
and $D^A$ is covariant derivative (for details of notation, see Appendix). We
use $\rho = -1$ for $AdS$ space, $\rho=0$ for flat space, and $\rho = 1$ for
$dS$ space. It is the use of operators  $\bar\Cbf_\st$, $\Cbf_\st$
\rf{0207200-07},\rf{0207200-08} that allows us to write down the concise
expression for operator $E$ in \rf{0207200-05}. For massless field in
$(A)dS$, $\ebf_1=\bar\ebf_1=0$, our operator $\bar\Cbf_\st$ \rf{0207200-07}
coincides with the standard de Donder operator in $(A)dS$ background.

Lagrangian \rf{0207200-04} is invariant under gauge transformation
\beq
\label{gautraarbspi01bf}
&& \delta \Phik = \Gbf \Xik\,,\qquad \Gbf \equiv \alphabf \Dbf - \ebf_1  -
\frac{\alphabf^2}{2s+d-5-2N_\zeta}\bar\ebf_1\,,
\\[3pt]
&& \Xik \equiv \sum_{s'=0}^{s-1}  \frac{\zeta^{s-1-s'} \alpha^{A_1} \ldots
\alpha^{A_{s'}} }{s'!\sqrt{(s - 1 - s')!}}
\, \Xi^{A_1\ldots A_{s'}} |0\rangle\,,
\eeq
where gauge transformation parameters $\Xi^{A_1\ldots A_{s'}}$  are
traceless, $\Xi^{AAA_3\ldots A_{s'}}=0$, i.e., $\bar\alphabf^2\Xik=0$. Also
we note that Lagrangian \rf{0207200-04} can alternatively be represented as
\beq
\label{02072009-12} \LL & = &  \half e \langle \Phi| (1-\frac{1}{4}\alphabf^2
\bar\alphabf^2)\EE \Phik \,,
\\[3pt]
&& \label{02072009-13} \EE  = \Box_{_{\rm (A)dS}} + \mbf_1 + \rho
\alphabf^2\bar\alphabf^2 - \Gbf\bar\Cbf_\st\,.
\eeq
Using \rf{09072009-01} and denoting eigenvalues of $N_\zeta$ by $k$, we find
the critical values of the mass parameter, $m_k^2 =\rho k(2s+d-4-k)$,
$k=0,1,\ldots, s-1$. The case $k=0$ corresponds to massless field, while
$k=1,2,\ldots,s-1$ correspond to the partial massless fields
\cite{Deser:1983mm,Deser:2003gw} (see also
\cite{Zinoviev:2001dt,Metsaev:2006zy,Skvortsov:2006at}).

{\it Standard de Donder gauge}. We proceed with discussion of standard de
Donder gauge for $(A)dS$ massive field. Representation for Lagrangian given
in \rf{0207200-04},\rf{0207200-05} is well adopted for this purpose. This is
to say that use of the following {\it de Donder} gauge-fixing term
\beq
\label{22082008-01xx} \LL_{{\rm g.fix}} & = & \half e \Phibr \Cbf_\st
\bar\Cbf_\st \Phik \,,
\eeq
leads to de Donder gauge fixed Lagrangian $\LL_\total$, $\LL_\total \equiv
\LL + \LL_{{\rm g.fix}}$,
\beq \label{02072009-12x}
\LL_\total & = & \half e \Phibr (1-\frac{1}{4}\alphabf^2 \bar\alphabf^2)
\EE_\total \Phik \,,
\\[5pt]
\label{02072009-13x} && \EE_\total =   \Box_{_{\rm (A)dS}} + \mbf_1 + \rho
\alphabf^2\bar\alphabf^2\,.
\eeq
Note that Lagrangian \rf{02072009-12} leads to the following gauge invariant
equations of motion $\EE\Phik$=0, where $\EE$ is given in \rf{02072009-13}.
It easy to see that imposing standard de Donder gauge $\bar\Cbf_\st\Phik=0$
we obtain gauge fixed equations of motion $\EE_\total\Phik=0$, where
$\EE_\total$ is given in \rf{02072009-13x}.

{\it Modified de Donder gauge}. We now discuss modified de Donder gauge. From
now on we consider fields in $AdS$, i.e. we set $\rho=-1$, and use Poincar\'e
parametrization of $AdS$. The modified de Donder gauge fixing is defined to
be
\beq
&& \LL_{{\rm g.fix}} = \half e \Phibr \Cbf \bar\Cbf \Phik \,,
\\[5pt]
\label{defgaucon003} &&  \Cbf \equiv \Cbf_\st - 2 \Cbf_\perp^z \,,
\hspace{2.5cm}
\bar\Cbf \equiv \bar\Cbf_\st + 2\bar\Cbf_\perp^z \,,
\\[5pt]
\label{defgaucon005} && \Cbf_\perp^z \equiv \alpha^z  - \half \alphabf^2
\bar\alpha^z \,,
\hspace{2cm}
\bar\Cbf_\perp^z \equiv \bar \alpha^z -\half \alpha^z \bar\alphabf^2 \,.
\eeq
We now make sure that gauge fixed Lagrangian $\LL_\total$, $\LL_\total \equiv
\LL + \LL_{{\rm g.fix}}$, takes the form
\beq
\LL_\total  & = &   \half e
\Phibr(1-\frac{1}{4}\alphabf^2\bar\alphabf^2)\EE_\total \Phik \,,
\\[5pt]
\label{02072009-11} \EE_\total  & = &  \Box_{_{0\,\rm AdS}} - m^2 - \alpha^2
\bar\alpha^z\bar\alpha^z - (s+\frac{d-4}{2} - N_z)^2
\nonumber\\[3pt]
& - & N_\zeta (2s+d-2 + 2N_z - N_\zeta) + \frac{d^2}{4} + 2\widetilde\Ibf^z
\bar\ebf_1 - 2\ebf_1\bar\alpha^z\,,
\\
&& \Box_{_{0\,\rm AdS}} \equiv z^2(\Box+\partial_z^2) +(1-d)z\partial_z,
\qquad \widetilde\Ibf^z = \alpha^z -\alphabf^2
\frac{1}{2N_\alphabf+d-1}\bar\alpha^z\,.\qquad
\eeq

We proceed with discussion of gauge-fixing procedure at the level of
equations of motion. To this end we note that gauge invariant Lagrangian
\rf{02072009-12} leads to the following equations of motion:
\be \label{28082008-01} \EE \Phik=0 \,,\ee
where $\EE$ is given in \rf{02072009-13}. We now define modified de Donder
gauge conditions as
\be
\label{20082008-01}  \bar\Cbf|\Phi\rangle  =  0 \,, \ee
where $\bar\Cbf$ is given in \rf{defgaucon003}. Using \rf{20082008-01} in
\rf{28082008-01} we get gauge fixed equations of motion
\be \label{20082008-02} \EE_\total \Phik=0 \,,\ee
where $\EE_\total$ is given in \rf{02072009-11}. We note that, because of
$\Cbf_\perp^z$- and $\bar{\Cbf}_\perp^z$-terms, the modified de Donder gauge
breaks some of the $so(d,2)$ symmetries. In the conformal algebra
nomenclature, these broken symmetries correspond to broken conformal boost
$K^a$-symmetries.

From $\EE_\total$ \rf{02072009-11}, we see that, because of terms like
$\alpha^2\bar\alpha^z\bar\alpha^z$, $\widetilde\Ibf^z\bar\ebf_1$, and
$\ebf_1\bar\alpha^z$ the modified de Donder gauge itself does not lead
automatically to decoupled gauge fixed equations for the ket-vector $\Phik$.
It turns out that in order to obtain decoupled gauge fixed equations of
motion we should introduce our fields in \rf{18052008-02}. We remind that
$\Phik$ is a double-traceless field \rf{froketvectcon} of the $so(d,1)$
algebra, while $\phik$ describes double-traceless fields \rf{phidoutracona01}
of the $so(d-1,1)$ algebra. This is to say that to get decoupled equations of
motion we have to make transformation from the $so(d,1)$ ket-vector $\Phik$
to $so(d-1,1)$ ket-vector $\phik$. We find a transformation from the
ket-vector $\Phik$ to our ket-vector $\phik$ and the corresponding inverse
transformation,
\beq
&& \label{05012008} \phik  = z^{\frac{1-d}{2}} V^\dagger \NN
\Pi^{\phi\Phi}\Phik\,,
\\[5pt]
&& \label{05012008-20}  \Phik  = z^{\frac{d-1}{2}} \Pi^{\Phi\phi} \NN  V
\phik\,, \eeq
where $V$ is unitary operator, $V^\dagger V =1$, and we introduce the
$z$-factor in r.h.s. of \rf{05012008} to obtain canonically normalized
ket-vector $\phik$. Operators $\Pi^{\Phi\phi}$, $\Pi^{\phi\Phi}$, $\NN$, and
$V$ are defined in the Appendix.

We now ready to compare modified de Donder gauges for $\phik$ \rf{080405-01}
and $\Phik$ \rf{20082008-01}. Inserting \rf{05012008-20} in \rf{20082008-01},
we make sure that modified de Donder gauge for $\Phik$ \rf{20082008-01}
amounts to one for $\phik$ \rf{080405-01} i.e., modified de Donder gauges for
$\phik$ \rf{080405-01} and $\Phik$ \rf{20082008-01} match. Also one can make
sure that gauge invariant Lagrangian for $\Phik$ \rf{0207200-04} and the one
for $\phik$ \rf{spi2lag01} match.

We now compare gauge transformation of $\phik$ \rf{gautraarbspi01} and gauge
transformation of $\Phik$ given in \rf{gautraarbspi01bf}. To this end we note
that gauge transformation parameters $\xik$ and $\Xik$ are related as
\be \label{22082008-02} \xik  =  z^{\frac{3-d}{2}} V^\dagger \NN'
\Pi_\alpha^\smpone \Xik \,,\quad
\qquad
\Xik  =  z^{\frac{d-3}{2}}  \Pi_{\alphabf}^\smpone \NN' V \xik \,,\qquad \ee
\be \NN' \equiv \NN|_{N_\alpha\rightarrow N_\alpha+1}\,, \ee
where $\Pi_\alpha^\smpone$, $\Pi_{\alphabf}^\smpone$ are defined in the
Appendix. Using \rf{05012008-20},\rf{22082008-02}, we make sure that gauge
transformations \rf{gautraarbspi01} and \rf{gautraarbspi01bf} match.

Finally we compare realization of $so(d,2)$ symmetries on the ket-vectors
$\phik$ and $\Phik$. To this end we note that on space of $\Phik$ realization
of the $so(d,2)$ algebra transformations takes the form
\beq
\label{conalggenlis01fr} && \delta_{P^a}\Phik  = \partial^a \Phik \,,
\qquad
\ \ \ \ \ \delta_{J^{ab}}\Phik = (x^a\partial^b -  x^b\partial^a +
M^{ab})\Phik\,,
\\[3pt]
\label{conalggenlis03fr} && \delta_D\Phik = x^B\partial^B \Phik\,,
\
\qquad \label{conalggenlis04fr} \delta_{K^a} \Phik = (-\frac{1}{2}x^Bx^B
\partial^a + x^a x^B\partial^B + M^{aB}x^B)\Phik \,,
\eeq
where $x^Bx^B=x^bx^b + z^2$, $x^B\partial^B = x^b\partial^b + z\partial_z$,
$M^{aB}x^B = M^{ab}x^b - M^{za}z$. Comparing \rf{conalggenlis01} and
\rf{conalggenlis01fr}, we see that the realizations of Poincar\'e symmetries
on $\phik$ and $\Phik$ match from the very beginning. Taking into account
$z$-factor in \rf{05012008-20}, it is easily seen that $D$-transformations
for $\phik$ \rf{conalggenlis03} and $\Phik$ \rf{conalggenlis03fr} match.
After this, we make sure that realizations of the operator $K^a$ on $\phik$
\rf{conalggenlis04} and on $\Phik$ \rf{conalggenlis04fr} also match.

{\it Light-cone Lagrangian}. Using gauge invariant action \rf{spi2lag01} and
imposing light-cone gauge, we find the light-cone Lagrangian
\be
 \LL_{{\rm l.c.}} = \sum_{s'=0}^s\,\, \sum_{\lambda = [s-s']_2}
\frac{1}{2 s'!}\, \phi_\lambda^{i_1\ldots i_{s'}} \Box_{\kappa+\lambda}^{}
\phi_\lambda^{i_1\ldots i_{s'}}\,,
\ee
where $\Box_{\kappa+\lambda}$ is defined in \rf{02072009-03} and transverse
indices take values $i=1,2,\ldots,d-2$. As usually, the light-cone fields
$\phi_\lambda^{i_1 \ldots i_{s'}}$ are traceless, $\phi_\lambda^{iii_3\ldots
i_{s'}}=0$.

To summarize, using the Poincar\'e parametrization of $AdS$ space, we have
developed the CFT adapted formulation of massive totally symmetric arbitrary
spin $AdS$ field. In recent years, mixed symmetry fields have attracted
considerable interest (see e.g.
Refs.\cite{Moshin:2007jt}-\cite{Alkalaev:2008gi}). We think that
generalization of our approach to the case of mixed symmetry massless and
massive $AdS$ fields might be useful for study of dynamical aspects of such
fields. In this respect, it would be interesting to find generalization of
the modified de Donder gauge to the case of mixed symmetry fields.

\setcounter{section}{0} \setcounter{subsection}{0}
\appendix{ Notation }

Vector indices of the $so(d-1,1)$ algebra  take the values $a,b,c=0,1,\ldots
,d-1$, while vector indices of the $so(d,1)$ algebra take the values
$A,B,C=0,1,\ldots ,d-1,d$. We use mostly positive flat metric tensors
$\eta^{ab}$, $\eta^{AB}$. To simplify our expressions we drop $\eta_{ab}$,
$\eta_{AB}$ in the respective scalar products, i.e., we use $X^a Y^a \equiv
\eta_{ab}X^a Y^b$, $X^A Y^A \equiv \eta_{AB}X^A Y^B$,
$\eta^{AB}=(\eta^{ab},1)$. Using the identification $X^d \equiv X^z$ gives
the following decomposition of the $so(d,1)$ algebra vector: $X^A=X^a,X^z$.
This implies $X^AY^A = X^aY^a + X^zY^z$.

We use the creation operators $\alpha^a$, $\alpha^z$, $\zeta$ and the
respective annihilation operators $\bar{\alpha}^a$, $\bar{\alpha}^z$,
$\bar\zeta$
\be
[\bar{\alpha}^a,\alpha^b]=\eta^{ab}\,, \quad
[\bar\alpha^z,\alpha^z]=1\,,\quad [\bar\zeta,\zeta]=1\,,
\quad
\bar\alpha^a |0\rangle = 0\,,\quad  \bar\alpha^z |0\rangle = 0\,, \quad
\bar\zeta |0\rangle = 0\,.\ee
These operators are referred to as oscillators in this paper. The oscillators
$\alpha^a$, $\bar\alpha^a$ and $\alpha^z$, $\zeta$, $\bar\alpha^z$,
$\bar\zeta$, transform in the respective vector and scalar representations of
the $so(d-1,1)$ algebra and satisfy the hermitian conjugation rules,
$\alpha^{a\dagger} = \bar\alpha^a$, $\alpha^{z\dagger} = \bar\alpha^z$,
$\zeta^{\dagger} = \bar\zeta$. Oscillators  $\alpha^a$, $\alpha^z$ and
$\bar\alpha^a$, $\bar\alpha^z$ are collected into the respective $so(d,1)$
algebra oscillators $\alpha^A =\alpha^a,\alpha^z$ and $\bar\alpha^A
=\bar\alpha^a,\bar\alpha^z$.

$x^A = x^a,z $ denote coordinates in $d+1$-dimensional $AdS_{d+1}$ space,
\be \label{speech01}
ds^2 = \frac{1}{z^2}(dx^a dx^a + dz dz)\,,
\ee
while $\partial_A=\partial_a,\partial_z$ denote the respective derivatives,
$\partial_a \equiv \partial / \partial x^a$, $\partial_z \equiv
\partial / \partial z$. We use the notation $\Box=\partial^a\partial^a$,
$\alpha\partial =\alpha^a\partial^a$, $\bar\alpha\partial
=\bar\alpha^a\partial^a$, $\alpha^2 = \alpha^a\alpha^a$, $\bar\alpha^2 =
\bar\alpha^a\bar\alpha^a$, $N_\alpha =\alpha^a\bar\alpha^a$, $N_z
=\alpha^z\bar\alpha^z$, $N_\zeta = \zeta\bar\zeta$.  The covariant derivative
$D^A$ is given by $D^A = \eta^{AB}D_B$,
\be \label{vardef01}
D_A \equiv e_A^\mu D_\mu\,,  \qquad D_\mu \equiv
\partial_\mu
+\frac{1}{2}\omega_\mu^{AB}M^{AB}\,, \qquad M^{AB} \equiv \alpha^A
\bar\alpha^B - \alpha^B \bar\alpha^A\,, \ee
$\partial_\mu = \partial/\partial x^\mu$, where $e_A^\mu$ is inverse vielbein
of $AdS_{d+1}$ space, $D_\mu$ is the Lorentz covariant derivative and the
base manifold index takes values $\mu = 0,1,\ldots, d$. The $\omega_\mu^{AB}$
is the Lorentz connection of $AdS_{d+1}$ space, while $M^{AB}$ is a spin
operator of the Lorentz algebra $so(d,1)$. Note that $AdS_{d+1}$ coordinates
$x^\mu$ carrying the base manifold indices are identified with coordinates
$x^A$ carrying the flat vectors indices of the $so(d,1)$ algebra, i.e., we
assume $x^\mu = \delta_A^\mu x^A$, where $\delta_A^\mu$ is Kronecker delta
symbol. $AdS_{d+1}$ space contravariant tensor field, $\Phi^{\mu_1\ldots
\mu_s}$, is related with field carrying the flat indices, $\Phi^{A_1\ldots
A_s}$, in a standard way $\Phi^{A_1\ldots A_s} \equiv e_{\mu_1}^{A_1}\ldots
e_{\mu_s}^{A_s} \Phi^{\mu_1\ldots \mu_s}$. Helpful commutators involving the
covariant derivative $D^A$ and the oscillators $\alpha^A$, $\bar\alpha^A$ may
be found in Appendix in Ref.\cite{Metsaev:2008ks}.

For the Poincar\'e parametrization of $AdS_{d+1}$ space, vielbein
$e^A=e^A_\mu dx^\mu$, Lorentz connection $\omega^{AB}=\omega_\mu^{AB}dx^\mu$,
and $\omega^{ABC}=e^{A\mu}\omega_\mu^{BC}$ are given by
\be\label{eomcho01} e_\mu^A=\frac{1}{z}\delta^A_\mu\,,\qquad
\omega^{AB}_\mu=\frac{1}{z}(\delta^A_z\delta^B_\mu
-\delta^B_z\delta^A_\mu)\,,\qquad \omega^{ABC} = \eta^{AC}\delta_z^B -
\eta^{AB}\delta_z^C\,.\ee
With choice made in \rf{eomcho01}, the covariant derivative takes the form
$D^A= z \partial^A + M^{zA}$, $\partial^A=\eta^{AB}\partial_B$.

The operators $\Pi^{\phi\Phi}$, $\Pi^{\Phi\phi}$ used in the Section
\ref{comparissec} are defined by relations
\beq \label{05012008-05-01}
\Pi^{\phi\Phi}
& \equiv & \Pi_\alpha^\smpone + \alpha^2 \frac{1}{2(2N_\alpha + d)}
\Pi_\alpha^\smpone (\bar\alpha^2 + \frac{2N_\alpha +d}{2N_\alpha + d
-2}\bar\alpha^z\bar\alpha^z)\,,
\\[3pt]
\label{05012008-05-02} && \Pi_\alpha^\smpone \equiv
\Pi^\smpone(\alpha,0,N_\alpha,\bar\alpha,0,d)\,,
\eeq\beq
\label{05012008-05-03} && \NN \equiv \Bigl(\frac{2^{N_z} \Gamma( N_\alpha +
N_z + \frac{d-3}{2}) \Gamma( 2N_\alpha + d - 3)}{\Gamma(N_\alpha +
\frac{d-3}{2})\Gamma(2N_\alpha + N_z + d - 3)}\Bigr)^{1/2}\,,
\\[7pt]
\Pi^{\Phi\phi}
& \equiv & \Pi_{\alphabf}^\smpone +  \alphabf^2 \frac{1}{2(2 N_\alphabf + d
+1)}\Pi_{\alphabf}^\smpone (\bar\alpha^2 - \frac{2}{2 N_\alphabf + d
-1}\bar\alpha^z\bar\alpha^z) \,,
\\[3pt]
\label{05012008-05-02bf} && \Pi_{\alphabf}^\smpone \equiv
\Pi^\smpone(\alpha,\alpha^z,N_\alphabf,\bar\alpha,\bar\alpha^z,d+1)\,,
\eeq
\beq \label{pibasicdef01}
&& \Pi^\smpone(\alpha,\alpha^z, X,\bar\alpha,\bar\alpha^z, Y)
\equiv \sum_{n=0}^\infty (\alpha^2+\alpha^z\alpha^z)^n \frac{(-)^n \Gamma(X +
\frac{Y-2}{2} + n)}{4^n n! \Gamma(X + \frac{Y-2}{2} +
2n)}(\bar\alpha^2+\bar\alpha^z\bar\alpha^z)^n\,,\qquad
\eeq
where $N_\alphabf = N_\alpha + N_z$, $\alphabf^2= \alpha^2+\alpha^z\alpha^z$,
and $\Gamma$ is Euler gamma function. We note that the $\Pi_\alpha^\smpone$
in \rf{05012008-05-02} is obtained from \rf{pibasicdef01} by equating
$\alpha^z=\bar\alpha^z=0$, $X=N_\alpha$, $Y=d$, while
$\Pi_{\alphabf}^\smpone$ in \rf{05012008-05-02bf} is obtained from
\rf{pibasicdef01} by equating $X=N_\alphabf$, $Y=d+1$.

The operator $V$ used in the Section \ref{comparissec} is defined by
relations
\beq
V \ & = & \sum_{N=0}^s V^{(N)} \,,
\qquad
V^{(N)} =  \sum_{l,n=0,1,\ldots, N} V_{ln}^{(N)} \,,
\\[5pt]
V_{ln}^{(N)} & = & v_{ln}^{(N)} \alpha_z^{N-l}\zeta^l |0\rangle\langle 0|
\bar\alpha_z^{N-n} \bar\zeta^n\,,
\qquad
v_{nl}^{(N)}  =  \NN_{nl}^{(N)} X_{nl}^{(N)}\,,
\\[5pt]
\label{xxxdef-01} X_{nl}^{(N)} & = & \sum_{t=0}^l \frac{(-)^t(\kappa-N+l)_t}{
t! (l-t)! (\kappa+1)_t } X_{nt}\,,
\\[5pt]
\label{xxxdef-02} X_{nt} & = & \sum_{p=0}^{\min n, t} \frac{(n+1-p)_p
(t+1-p)_p (2s+d-2-n-p)_p}{p!(s+\frac{d-2}{2}-p)_p
(\kappa-s-\frac{d-4}{2})_p}\,,
\\[5pt]
\label{30032009-04}
\NN_{nl}^{(N)} & = &  \frac{(-)^n}{n!(N-n)!}
\Bigl(\frac{\kappa-N+2l}{\kappa}\Bigr)^{1/2}
\Bigl(\frac{2s+d-3-2n}{2s+d-3-n}\Bigr)^{1/2}
\nonumber\\[5pt]
& \times & \Bigl(\frac{(s+\frac{d-2}{2}-l)_l
(s+\frac{d-2}{2}-N+l)_{N-l}}{(2s+d-3-n-N)_N}\Bigr)^{1/2}
\nonumber\\[5pt]
& \times & \Bigl(\frac{(\kappa-s-\frac{d-4}{2})_n
(\kappa-s-\frac{d-4}{2})_l(\kappa+s+\frac{d-2}{2}-N+l)_{N-l}
(\kappa+1)_l}{(\kappa+s+\frac{d-2}{2}-n)_n (\kappa-N+l)_{N-l}}\Bigr)^{1/2}\,,
\eeq
where in \rf{xxxdef-01}-\rf{30032009-04} we use the notation $(a)_b$ for the
Pochhammer symbol, $(a)_b \equiv \frac{\Gamma(a+b)}{\Gamma(a)}$.

\end{document}